\documentclass[12pt,oneside,reqno]{amsart}
\usepackage[letterpaper,margin=1in]{geometry}                
\usepackage{amssymb,amsmath,amsfonts}
\usepackage{xcolor}
\usepackage{float}
\newcommand{\VVV}{{\mathsf{{V \hspace{-.195cm} I\,}}}}
\usepackage[colorlinks,
             linkcolor=blue!50!red,
             citecolor=blue!50!red,
             urlcolor=blue!50!red,
             pdfproducer={pdfLaTeX},
             pdfpagemode=None,
             bookmarksopen=true
             bookmarksnumbered=true]{hyperref}
\usepackage{graphicx}
\usepackage[utf8]{inputenc}

\title{Stokes Polytopes and Intersection Theory}

\author{Nikhil Kalyanapuram}
\date{\today}

\begin{document}

\maketitle

\begin{abstract}
Intersection numbers of Stokes polytopes living in complex projective space are computed using the techniques employed to find the inverse string KLT matrix elements in terms of intersection numbers of associahedra. To do this requires an appropriate convex realization of Stokes polytopes in $\mathbb{CP}^{n}$ loaded with suitable generalizations of the Koba-Nielsen factor. The procedure is carried out explicitly for the lower point cases and the prescription for the generic higher point cases is laid out as well. The intersection numbers are identified as scattering amplitudes corresponding to a theory the coupling constants of which are determined entirely in terms of the combinatorial weights of the Stokes polytopes. A parameter $\alpha'$ having units of length is used to define the intersection numbers in a manner that yields the amplitudes of $\phi^4$ theory to leading order when the limit of vanishing $\alpha'$ limit is taken. Most importantly, we contrast and compare this method of understanding quartic vertices with previous string-theoretic attempts to obtain quartic interaction amplitudes.
\end{abstract}

\section{Introduction}\label{section1}

In recent years, there has been a renewal of interest in the worldsheet approach to understanding scattering amplitudes for a large class of theories. A worldsheet picture of scattering amplitudes for a large class of field theories, including pure Yang-Mills and gravity was realized by the CHY formalism \cite{CHY1,CHY2,CHY3,CHY4}, in which the amplitudes are recast as integrals over the moduli space of punctured Riemann spheres $\mathcal{M}_{0,n}$. Unlike in the string theory case however, the full moduli space is not probed by these integrals. Rather, the integrand is localized over the Gross-Mende saddle points \cite{gross},

\begin{equation}
    \sum_{j}\frac{k_{i}\cdot k_{j}}{\sigma_{i} - \sigma_{j}}=0,
\end{equation}

where $i$ runs over all the particles in the theory and the $\sigma_{i}$ are the respective marked points on $\mathbb{CP}^{1}$. 

Of particular interest is the CHY representation of gauge theory and gravity amplitudes, which allows for a simple proof of the KLT relations\cite{KLT}. These relations are a prescription for computing gravity amplitudes given colour ordered amplitudes in Yang Mills theory. When the colour ordered $n$ point Yang Mills amplitudes $M_{n}[\alpha_{i}]$ and $M_{n}[\beta_{i}]$ are given for BCJ bases $\lbrace{\alpha_{i}\rbrace}_{i=1,\dots,(n-3)!}$ and $\lbrace{\beta_{i}\rbrace}_{i=1,\dots,(n-3)!}$\cite{BCJ}, the KLT kernel $K[\alpha_{i}|\beta_{j}]$ supplies the gravity amplitude via the convolution,

\begin{equation}
  M_{grav,n} =  \sum_{\lbrace{\alpha_{i}\rbrace},\lbrace{\beta_{i}\rbrace}} M_{n}[\alpha_{i}]K[\alpha_{i}|\beta_{j}]M_{n}[\beta_{j}].
\end{equation}

It should be noted that although it appears as though the above equation would have $(n-3)!\times(n-3)!$ terms, not all elements of the KLT matrix are nonzero, consequently reducing the number of terms considerably. Furthermore, different choices of the BCJ bases would yield different KLT relations, although these are equivalent to each other.

The KLT kernel however contains data in addition to this. Treated as a matrix, it was shown in \cite{CHY1} that the inverse $K^{-1}[\alpha_{i}|\beta_{j}]$ of the KLT kernel encodes as its matrix elements colour ordered amplitudes in the biadjoint theory (the reader is referred to \cite{CHY1} for a review of this theory).

The KLT kernel owes its origins to a computation in the context of string scattering amplitudes, in which it arises (with $\alpha'$ corrections) as the kernel fusing open string amplitudes of massless particles in a BCJ basis to obtain closed string amplitudes of massless particles. This has the field theory limiting case as the fusing of gauge theory amplitudes to obtain gravity scattering amplitudes. Now having understood the biadjoint theory as described by the inverse of the field theory KLT kernel, one may inquire as to the role of the inverse of the kernel that appears in string theory.

This question was addressed by Mizera in \cite{ref0}, in which this object was computed and shown to in a sense represent the amplitudes of an $\alpha'$ generalization of the biadjoint theory. Diagrammatic rules for computing these amplitudes were presented as well. 

The $\alpha'$ corrected biadjoint amplitudes were put in a larger, unified context in \cite{ref1,ref1-1}. In these, it was shown that a vast variety of amplitudes, included those arising in open string theory, the $\alpha'$ corrected biajoint amplitudes and CHY amplitudes arise as a consequence of the \emph{twisted intersection theory} of cycles and cocycles which are defined in terms of certain hyperplane arrangements in $\mathbb{CP}^{n}$. 

A twisted cycle is a region of integration with vanishing boundary that is dressed with a function that has branch cuts. Twisted cocyles are defined analogously as differential forms belonging to cohomology classes of the exterior derivative twisted by a connection. In \cite{ref2,ref3,ref4,ref4-1}, a systematic study was undertaken to study the theory of twisted cyles associated to a particular configuration of hyperplanes in $\mathbb{CP}^{n}$. There is an invariant pairing that can be defined given two twisted cycles, two twisted cocyles, or a twisted cycle and twisted cocycle, generically labelled as the twisted intersection pairing. 

In \cite{ref1,ref1-1}, it was established that this theory may be directly imported into the theory of scattering amplitudes. The moduli space $\mathcal{M}_{0,n}$ was realized as a hyperplane arrangement and it was observed that a polytope called the associahedron or Stasheff polytope, lives naturally inside the compactification of this space. Dressing the associahedron with the Koba Nielsen factor yields a twisted cycle, and it was shown that the Parke Taylor form arises as a twisted cocycle. Concordantly, the intersection of a cycle and cocycle yields the disk integral of Parke-Taylor form, or $Z$-theory amplitudes \cite{oliver1,oliver2,oliver3}, the intersection theory of two cycles yields the CHY amplitudes and most surprisingly, the intersection of twisted associahedra gave rise to the inverse KLT kernel.

Although the associahedron arises in this context as the compactification of the simplex living in $\mathcal{M}_{0,n}$, it was shown by Nima Arkani-Hamed et al. \cite{ref6} that the associahedron admits a natural embedding in kinematical space and is the amplitudehedron for the biadjoint scalar field theory. Combinatorially, the vertices of the associahedron encode the triangulations of an $n$-gon, or equivalently all planar Feynman diagrams for $n$ particles scattering in biadjoint scalar theory. This is indicated in the $\alpha'$ limit of the intersection numbers as well, in which the numbers localize on the vertices of the intersecting associahedra, as was established by Mizera in \cite{ref1}.

Along the lines of the associahedron program was developed a similar picture for $\phi^4$ theory in \cite{ref7}. Stokes polytopes realize the planar sector of $\phi^4$ amplitudes in the positive geometry program, much as the associahedron did so for the biadjoint theory. Importantly, the Stokes polytope for a given amplitude, say the $8$ point amplitude, is not unique. It was seen that at a given order, multiple Stokes polytopes contribute, and are weighted in the final amplitude by constants depending only on the combinatorial structure of the polytopes.

The primary task to be addressed in this note is then the following. The intersection theory of Stokes polytopes are studied and the effective field theory giving rise to such amplitudes is analyzed. The existence of a convex realization of the Stokes polytopes in projective space will be assumed, such that the facets are represented by hypersurfaces in the ambient space. Some simple examples of such embeddings will be supplied and understood to generalize suitably for higher dimensional versions. 

The organization of the paper is as follows. The techniques developed by Mizera in the context of the associahedron will be first employed to understand the intersection theory of Stokes polytopes. As will be elaborated upon further, one will have to include contributions of all the relevant Stokes polytopes for a given process. It will be attempted to obtain an effective Lagrangian for the amplitudes found, classified by powers of the coupling constant and it will be seen that as in the case of the weights of Stokes polytopes, the numerical coefficients depend purely on the combinatorial structure of the Stokes polytopes. Finally, the $\alpha'\rightarrow 0$ limit of the amplitudes will be identified with the regular amplitudes in $\phi^4$ theory. In doing so, we explore the worldsheet aspects of amplitudes in this theory. It is known that in this limit, intersection numbers localize on the boundaries of the moduli space and are given equivalently as cocycle intersection numbers. We will elaborate on this point and then draw attention to earlier attempts to extract $\phi^4$ amplitudes from worldsheet integrals in perturbative string theory, emphasizing how our approach improves upon such calculations.

\section{Intersection Theory of Twisted Cycles}\label{section2}
In this section, we attempt a short review of the intersection theory of what are known in the literature \cite{ref2,ref3,ref4,ref4-1} as twisted cycles. A comprehensive review of this fairly technical subject is not attempted, and for the proofs of the more involved statements, the reader is referred to the bibliography.

To begin, let us recall the definitions of the notions of twisted cycles. A twisted cycle is a cycle belonging to a twisted homology class of a manifold $X$. To understand this in the language of differential forms, we look instead at the twisted cohomology. Twisted cohomology is defined just as regular cohomology, excepting the fact that the differential operator receives a \emph{twist}, which is constructed as follows. Consider a hyperplane arrangement $\bigcup H_{i}$ in $\mathbb{CP}^{n}$, where the $H_{i}$ are hyperplanes given by the vanishing of linear relations $f_{i}$ among the homogeneous coordinates.

Delete now the hyperplanes from the complex projective space, to obtain what is called the configuration space $X  = \mathbb{CP}^n - \bigcup H_{i}$. The \emph{twist} is now a differential one form, that is supported on $X$, and suffers logarithmic divergences as the hyperplanes are approached,

\begin{equation}
    \omega  = \sum_{i} c_{i}d\log f_{i}.
\end{equation}

For the time being, the $c_i$'s are introduced as formal variables. Later, it will be seen that they can be identified with planar variables in scattering amplitudes. Generally, one would like that the sum over these constants vanishes, in order to avoid the appearance of residues at infinity. Generically in the cases we will be interested in later, this will often not be possible. In order to assure that there is no residue at infinity, one \emph{adds} additional hyperplanes at infinity, whose coefficients precisely cancel the ones coming from the hyperplanes not at infinity. This is often done by us implicitly, and is to be understood in that sense henceforth. 

Now, twisting the exterior derivative now amounts to introducing $\omega$ as a connection,

\begin{equation}
    \nabla_{\omega} = d + \omega \wedge.
\end{equation}

It is readily checked that this is a nilpotent operator, that squares to zero, hence defining a cohomology. Elements of the groups obtained thereby are referred to as twisted differential forms. Accordingly, $\omega$ is generally referred to as the \emph{twist}.

Twisted versions of the traditional homology group elements may also be defined. For our purposes, it is sufficient to look at the top dimensional case, for it is this class which will be relevant. Let us return to the hyperplane arrangement as previously defined. Now, we specialize to those integration domains whose only boundaries are those captured by the hyperplane arrangements. Generally, in all cases holding our interest, the hyperplanes will be chosen as boundaries of convex regions in $\mathbb{CP}^n$, and these regions will be the regions of integration. We denote such a convex region by $\Delta(H_{i})$. That this is a cycle is established by the vanishing of its boundaries in $X$, due to the removal of the hyperplanes. 

Loading this cycle is now the operation of dressing it with a specific branch of $\exp(\omega) = \prod_{i}f^{c_{i}}_{i}$. It is now possible to show that the intersection numbers obtained by integrating a twisted form $\phi_{\omega}$ over such a region, namely the number

\begin{equation}
    \int_{\Delta(H_{i})}\prod_{i}f^{c_{i}}_{i} \phi_{\omega},
\end{equation}

is a cohomological invariant. These are the so called twisted intersection numbers.

Now, instead of looking at the intersections of cycles and cocycles, we can define an intersection theory for two prescribed cycles as well. The twisting is done as follows. Let one of the cycles $\Delta_1$ be dressed using the twist $\omega$ and another cycle $\Delta_2$ be dressed using the twist $\overline{\omega}$. In this work, we will be interested in the self intersection case, so we take $\Delta_1 = \Delta_2$. As mentioned earlier, the cycles relevant to us will be defined as the interiors bounded by the chosen hyperplane arrangement. Consequently, the cycles are noncompact, thus requiring a regularization in order to admit intersection numbers. This procedure of regularization is described in detail in \cite{ref4-1}, to which the reader is directed. For the choice of cycles we have made in this work however, a combinatorial description of the intersection numbers is possible. This is done as follows.

Let us label the boundaries of the cycle in terms of the corresponding codimension. The codimensionality of the entire cycle would be $0$. Facets, namely the bounding hyperplanes would have codimension $1$. Working upwards, we will encounter edges of codimension $n-1$ and vertices of codimension $n$, where $n$ is the dimensionality of the ambient space. A codimension $k$ boundary is realized as the region of intersection of $k$ facets. Let these facets be $f_{i_1},\dots,f_{i_{k}}$. The twisted intersection number receives the following contribution from this boundary,

\begin{equation}
    \frac{1}{e^{2\pi i c_{i_{1}}} -1}\times\dots\times\frac{1}{e^{2\pi i c_{i_{k}}} -1}.
\end{equation}

A summation has to now be performed over all the boundaries, with the understanding that the contribution from the whole cycle, namely the barycentre is simply $1$. 

This prescription was used in \cite{ref1} to compute the intersections of associahedra, which naturally lie inside the compactified moduli space of $n$ puctured spheres $\mathcal{M}_{0,n}$. Since this moduli space can be realized as a specific hyperplane arrangement in $\mathbb{CP}^{n-3}$, the tools of twisted intersection theory can be applied. When the associahedra are loaded with Koba-Neilsen factors and the branch of these factors is chosen in terms of a BCJ basis, the intersection numbers so obtained are precisely the matrix element of the inverse KLT kernel in string theory.

Our main task in this present work will be to extend this analysis to the case of Stokes polytopes and understand the structure of these intersection numbers. It is a well known fact that the vertices of the associahedron $K_{n-3}$ encode complete triangulations of an $n$ gon, which has the interpretation of noting planar amplitudes in cubic scalar theory. Consequently, in the limit $\alpha'\rightarrow 0$ where only the vertices contribute in the intersection numbers, the inverse KLT kernel elements reduce to biadjoint scalar amplitudes. It was shown in \cite{ref1-1} that this limit may also be computed as the intersection numbers of cocycles, which localized on the boundaries of the moduli space and are given essentially by the CHY formula. This suggests that a similar analysis applied to Stokes polytopes should yield CHY type formulae for the computation of amplitudes in $\phi^4$ theory, which will be the topic of concern for us in section \ref{sec6}.

\section{Examples of Intersection Numbers for Stokes Polytopes}

In this section, attention will be turned to computing the intersection numbers of Stokes polytopes. Now Stokes polytopes admit convex realizations that allow their descriptions in terms of vertices, edges and so on. Before explicitly presenting the results of the computations, it is worth noting how the intersection numbers schematically present themselves.

The Stokes polytope in one dimension is just the associahedron. In accordance with this, the intersection numbers come from vertices and the barycentre of the associahedron, which is simply the center of a line. Thus, we receive a constant, namely $1$ from the vertices, and terms $\frac{1}{e^{i\alpha' X_{H}} - 1}$ from the vertices, where $X_{H}$ is the planar variable corresponding to the vertex. We emphasize at this point that the variable $\alpha'$ is chosen simply as a regulating parameter to ensure that the factor in the exponential is dimensionless and is not intended to bear any suggestive relationship with the string tension. Since it is ultimately the limit of vanishing $\alpha'$ with will be our concern, we hope that this will not cause any confusion.

Now for the $8$ particle case, the Stokes polytope is two dimensional. As a result, contributions from the vertices, edges and the barycentre are obtained. The barycentre as usual gives unity, while the edges expressions of the form $\frac{1}{e^{i\alpha' X_{H}} - 1}$, where as usual $X_{H}$ is the planar variable corresponding to a facet. Differently from the four point case, the vertices are now associated with the points of intersection of two facets, and now contribute factors $\frac{1}{\left(e^{i\alpha' X_{H_{1}}} -1\right)\left(e^{i\alpha' X_{H_{2}}} -1\right)}$, where $H_{1}$ and $H_{2}$ are the facets giving rise to the vertex in question and the planar variables are labelled in accordance with this fact. We will illustrate this general rule with a specific example.

Generally, the intersection numbers are calculated using analogous rules for higher order Stokes polytopes. Terms are organized in terms of the contributing boundaries of the Stokes polytopes. The most singular terms arise out of the vertices, followed by successively less singular terms until a pure constant is supplied by the barycentre. The only subtlety to be taken note of is the non unique character of the Stokes polytopes. The fact that there is more than one Stokes polytope in general in a given dimension requires us to sum over all such polytopes. The weights are provided by the requirement that the terms arising out of vertec contributions must have unit residue. This is the same as requiring the $\alpha'\rightarrow 0$ limit return the field theory expression upto a divergence.

A word on notation may be said at this point. Stokes polytopes will be constructed as convex hulls of certain hyperplane arrangements. Since they depend on the choice of quadrangulation $Q$, an $n$ dimensional Stokes polytope coming from quadrangulation $Q$ will be denoted by $\mathcal{S}_{n,Q}$. These are topological objects which will now be loaded with specific multivalued functions vanishing on the boundaries of these objects. These functions will be denoted by $\mathrm{Stokes}_{n,Q}(z_i)$, where the dependence on the inhomogeneneous coordinates is indicated. The task to now be performed is the computation of the intersection numbers which will be denoted by $\langle{\mathcal{S}_{n,Q}\otimes\mathrm{Stokes}_{n,Q}(z_i),\mathcal{S}_{n,Q}\otimes\mathrm{Stokes}_{n,Q}(z_i)\rangle}$

\subsection{Intersection Theory for the $6$ - Particle Amplitude}\label{section3}
This section will be concerned with the computation of the intersection theory of the six point Stokes polytopes. In order to carry out these, the Stokes polytopes for a quadrangulated hexagon must be found. The quadrangulations correspond to various factorization channels, in a manner entirely analogous to the role played by triangulations for the biadjoint amplitudes. 

It is immediately observed that there is a single topologically invariant quadrangulation, namely the one obtained by introducing a diagonal between vertices $1$ and $4$, corresponding to the planar variable $X_{14}$, which labels the momentum transfer across this channel. This is due to the fact that any other quadrangulation is obtained by a rotation of the foregoing.

Now, the diagonal $X_{36}$ is $Q$ compatible with $X_{14}$, thus defining the Stokes polytope. Such a Stokes polytope takes the form of a line, whose vertices are labelled by these diagonals. The intersection theory of this is now identical to that of a one dimensional associahedron, and gives the intersection number as,

\begin{equation}\label{eq3.1}
\begin{aligned}
  &\langle{(-1,1)\otimes\mathrm{Stokes}_{1,(14)}(z),(-1,1)\otimes\mathrm{Stokes}_{1,(14)}(z)\rangle}\\& =  1 + \frac{1}{e^{2\pi i\alpha' X_{14}}-1} + \frac{1}{ e^{2\pi i\alpha' X_{36}}-1}. 
\end{aligned}
\end{equation}

Let us now understand how the above expression is obtained as the intersection number of a twisted cycle.

The Stokes polytope in one dimension is simply a line, which is the one dimensional associahedron. Consequently, it is possible to import that analysis used in the case of the latter to obtain the intersection number. The one dimensional associahedron lives in the Deligne-Mumford compactification of the moduli space $\mathcal{M}_{0,4}$ of Riemann spheres with four punctures - configuration space $\mathbb{CP}^{1}-\lbrace{-1,1,\infty\rbrace}$\footnote{The unusual choice of removing $-1$ and $1$ is motivated by the fact that later in this work we will observe how Stokes polytopes are realized as convex polytopes, which can be defined as the interior of certain hyperplane arrangements, which in the case of the one dimensional Stokes polytope will simply be the two hyperplanes $z=\pm1$ in $\mathbb{CP}^1$. This should not cause any confusion however, since conformal transformations may be used readily to bring these into the canonical points $0$ and $1$.\label{ftn1}}. It takes the form of the interval $(-1,1) = \mathcal{S}_{1,(14)}$, and carries the function,

\begin{equation}
  \mathrm{Stokes}_{1,(14)}(z) =  (z+1)^{\alpha'X_{14}}(1-z)^{\alpha' X_{36}},
\end{equation}

where the notation is used to indicate the one dimensional Stokes polytope corresponding to the dissection $X_{14}$\footnote{It may be noticed that this multivalued function bears some resemblance to the Koba-Nielsen factor from disk amplitudes. Indeed, there is the question of whether or not there is an uplift of such functions which may then admit a more symmetric representation. Geometrically, this corresponds to viewing Stokes polytopes as embeddings in associahedra \cite{alok, alok2}, which is an interesting problem in its own regard. I thank Oliver Schlotterer for bringing this question to my attention.}.

The intersection theory of such cycles was studied in \cite{ref2,ref3,ref4,ref4-1} and applied to the case of associahedra in \cite{ref1}. The reader may notice that the cycles are noncompact, implying that a regularization is required in order to define the intersection numbers. The reader is pointed to these references for further details regarding such subtleties. The intersection number $\langle{(-1,1)\otimes\mathrm{Stokes}_{1,(14)}(z),(-1,1)\otimes\mathrm{Stokes}_{1,(14)}(z)\rangle}$ is precisely equation \ref{eq3.1} when the regularization is correctly carried out.

In order to obtain the full intersection theory amplitude, one has to include contributions from all the Stokes polytopes corresponding to all possible quadrangulations weighted appropriately as explained in \cite{ref7}. As indicated already, this is done by simply permuting the indices on the planar variables, on account of the fact that a single topologically distinct quadrangulation exists. The weights for all the Stokes polytopes at this level are $\frac{1}{2}$. Summing over all the Stokes polytopes weighted as such give,

\begin{equation}\label{eq3.2}
    \frac{3}{2} + \frac{1}{e^{2\pi i\alpha' X_{14}} - 1} +  \frac{1}{e^{2\pi i\alpha' X_{25}}-1} +  \frac{1}{ e^{2\pi i\alpha' X_{36}} -1}.
\end{equation}

This can be written more simply as, 

\begin{equation}
    \frac{i}{2}\left(\frac{1}{\tan{2\pi\alpha'X_{14}}} + \frac{1}{\tan{2\pi\alpha'X_{25}}} + \frac{1}{\tan{2\pi\alpha'X_{36}}}\right).
\end{equation}

While this is not the same as the inverse KLT kernel for four particles, it bears a striking resemblance, predominantly due to the fact that the polytope controlling the amplitudes for four particle scattering in $\phi^3$ theory and the polytope controlling the amplitudes for six particle scattering amplitudes in $\phi^4$ theory are one dimensional associahedra.

Now, the effective action corresponding to the above scattering amplitude may be recorded. The propagator of this theory is,

\begin{equation}
    \frac{1}{e^{-2\pi i\alpha'\square}-1}.
\end{equation}

Inspecting now equation \ref{eq3.2}, the amplitude decomposes into two parts. One is a six point vertex and the other is made up of two four point vertices and propagators. Both are proportional formally to $\lambda^2$, where $\lambda$ is the four point coupling in the field theory. Thus, two terms are obtained in the interacting part of effective Lagrangian,

\begin{equation}
    \mathcal{L}^{eff}_{I} = \lambda\varphi^4 + \frac{3}{2}\lambda^2\varphi^6 + O(\varphi^{8}).
\end{equation}

Having computed the intersection numbers for the simplest case, now we have to perform similar computations for the higher point cases. Before moving on, a general feature of these intersection numbers can be noted. It can be seen, almost by inspection that the $n$ point contact term at any order is simply given by a sum over all the weights. 

\subsection{Intersection Theory for the $8$ - Particle Amplitude}\label{sec3.2}
In this section, our main concern will be the computation of intersection numbers that are obtained from Stokes polytopes corresponding to the eight point scattering amplitudes in $\phi^4$ theory. This is the case that first differs from the treatment of intersection numbers in \cite{ref1}, since Stokes polytopes in two dimensions are not unique. The Stokes polytopes in two dimensions were first computed in the context of scattering amplitudes in \cite{ref7}, and are of two varieties. Combinatorially, these polytopes are topologically squares or pentagons. Let us review the construction of these polytopes before proceeding to the calculation of the intersection numbers. The reader may refer to \cite{ref7} for the original treatment, which we recall now.

An $8$ particle scattering process in the planar sector is schematically represented by an octagon whose vertices are labelled as $1,2,\dots,8$ and the $i^{th}$ particle is viewed as entering via the edge $(i,i+1)$. With this construction, one of the contributing Feynman diagrams is a quadrangulation of this octagon. Consider first the quadrangulation $\lbrace{(14),(58)\rbrace}$, which indicates a joining of vertices $1$ and $4$ and a joining of vertices $5$ and $8$. Now, in order to obtain the Stokes polytope associated to this quadrangulation, which we will denote by $Q_{(14),(58)}$, we must look at all quadrangulations and subject them to the condition of $Q$ compatibility, where $Q$ is the quadrangulation $\lbrace{(14),(58)\rbrace}$. Those that pass this test collectively denote the vertices of the Stokes polytope. For the quadrangulation under consideration, these vertices are,

\begin{equation}\label{eq3.7}
    \lbrace{\lbrace{(14),(58)\rbrace},\lbrace{(14),(47)\rbrace},\lbrace{(38),(58)\rbrace},\lbrace{(38),(47)\rbrace}\rbrace}.
\end{equation}

Now this is realized as a polytope in the following manner. The vertex $\lbrace{(14),(58)\rbrace}$ is found at the intersection of two \emph{facets}, which in this case are of codimension $1$ in an ambient space of dimension $2$, which are given combinatorially as $(14)$ and $(58)$. The other vertices are obtained similarly at intersections of two out of the facets $(14)$, $(47)$, $(38)$ and $(58)$. Accordingly, the Stokes polytope in this case is combinatorially a square, with four facets and four vertices.

\begin{figure}[h]
    \centering
    \includegraphics[width=0.5\textwidth]{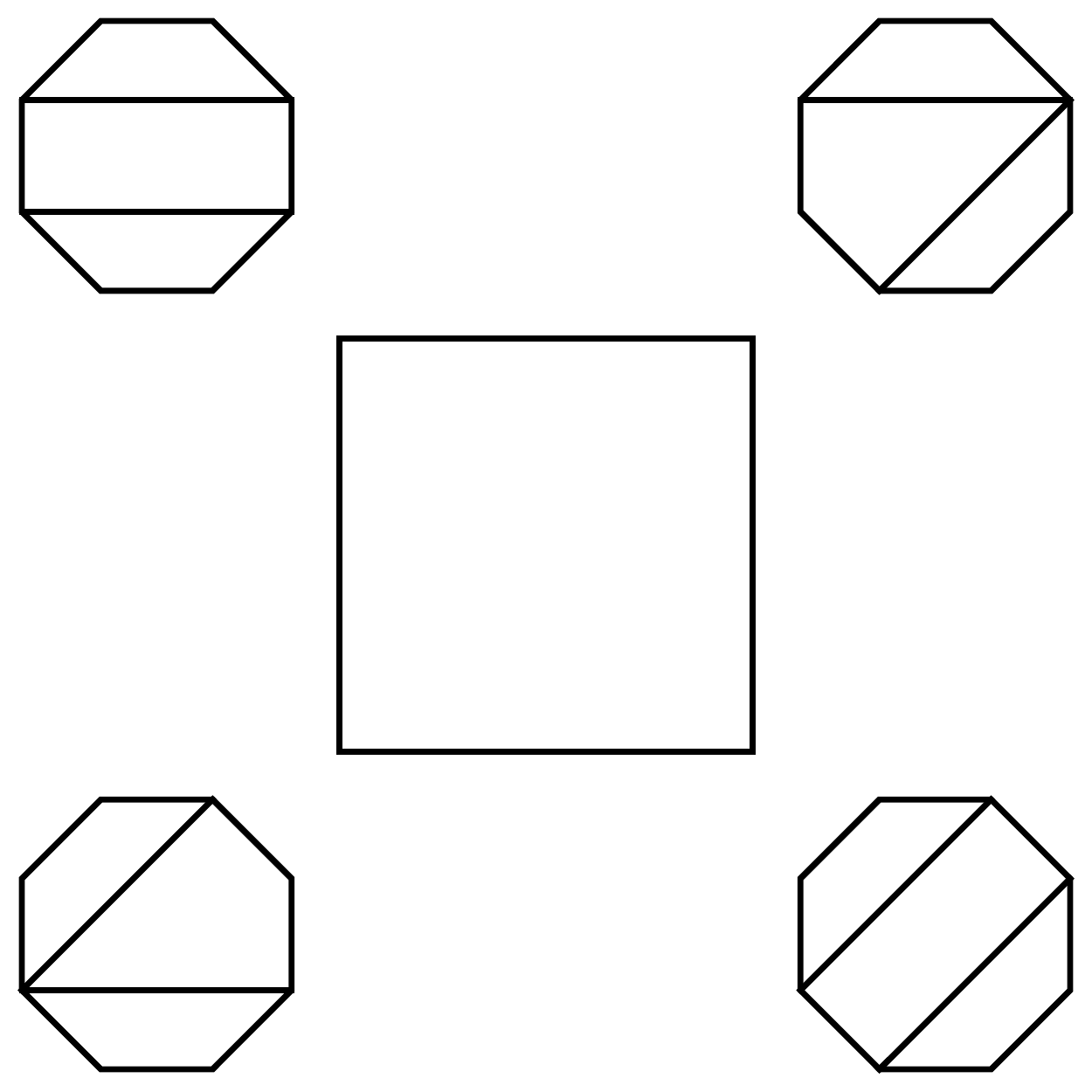}
    \caption{The Stokes polytope given by equation \ref{eq3.7}. The upper left corner denotes the reference dissection.}
    \label{fig:mesh1}
\end{figure}

In a later section, we will explain in detail how Stokes polytopes arise as convex polytopes by defining the embedding of their facets as hyperplanes in $\mathbb{R}^n$. Complexifying coordinates then realizes these polytopes in complex space, to which then the standard techniques of twisted intersection theory can be applied. For the time being, we assume that these hyperplane arrangements are understood, and simply denote the hyperplanes as $H_I$. For the case at hand, there are four facets, which collectively represent the four choices of $I$.
  
The cycle to be considered now is the interior region bounded by the hyperplanes $H_i = 0$. The hyperplanes are arranged such that $H_{(14)}$ and $H_{(58)}$ intersect, as do $H_{(58)}$ and $H_{(38)}$, $H_{(38)}$ and $H_{(47)}$ and $H_{(47)}$ and $H_{(14)}$, thus forming a quadrilateral. The subscripts have been used to denote the facets of the Stokes polytopes represented by the respective hyperplane. The interior so obtained is the Stokes polytope $\mathcal{S}_{2,\lbrace{(14,58)\rbrace}}$. In corresponding notation, the cycle is to be now \emph{loaded}, namely assigned the following function,

\begin{equation}
    \mathrm{Stokes}_{2, (14,58)}(z_1,z_2) = H_{1}^{\alpha' X_{14}}H_{2}^{\alpha' X_{58}}H_{3}^{\alpha'X_{38}}H^{\alpha'X_{47}}_{4}.
\end{equation}

Dealing with a cycle in two dimensions now means that we have to consider three classes of contributions, namely those coming from the barycentre of the Stokes polytope, from the facets and from the vertices. We will describe these one by one. Consider first the contribution from the barycentre. Barycentres just contribute $1$ to the intersection number. 

Before computing the contributions of the facets, let us note that the barycentre of every Stokes polytope will simply give a factor $1$. This means that again, the sum over all Stokes polytope intersection numbers will obtain contributions from the barycentres equal to a sum over all the weights of the Stokes polytopes. It was shown in \cite{ref7} that there are twelve Stokes polytopes, four of which are squares, weighted by factors of $\frac{2}{6}$ and the remaining eight are pentagons weighted by factors of $\frac{1}{6}$. In accordance with these, we see that the total barycentre contribution from the Stokes polytope intersection numbers for eight particle scattering is $\frac{16}{6}$. In the effective action, this will give rise to an eight point contact term with coupling constant $\frac{16}{6}\lambda^3$.

Let us now return to the remainder of the computation of the Stokes polytope $Q_{(14),(58)}$ with itself. We now have to consider contributions that come from the faces and the vertices. The faces each contribute a term,

\begin{equation}
    \frac{1}{e^{2\pi i\alpha'X_{ij}}-1},
\end{equation}

where the subscript $(ij)$ is understood to denote the corresponding facet. Each vertex contributes a factor that is a product of two terms like the one above, coming from the two facets that intersect to give a vertex. In all, we have the following intersection number,

\begin{equation}
    \begin{aligned}
    &\langle{\mathcal{S}_{2,(14,58)}\otimes\mathrm{Stokes}_{2, (14,58)}(z_1,z_2),\mathcal{S}_{2,(14,58)}\otimes\mathrm{Stokes}_{2, (14,58)}(z_1,z_2)\rangle}=\\
       & 1 + \frac{1}{e^{2\pi i\alpha'X_{14}}-1} + \frac{1}{e^{2\pi i\alpha'X_{38}}-1}  + \frac{1}{e^{2\pi i\alpha'X_{47}}-1} + \frac{1}{e^{2\pi i\alpha'X_{58}}-1}\\
       &+ \frac{1}{e^{2\pi i\alpha'X_{14}}-1}\frac{1}{e^{2\pi i\alpha'X_{58}}-1}
         + \frac{1}{e^{2\pi i\alpha'X_{14}}-1}\frac{1}{e^{2\pi i\alpha'X_{47}}-1}\\& + \frac{1}{e^{2\pi i\alpha'X_{38}}-1}\frac{1}{e^{2\pi i\alpha'X_{58}}-1}
         + \frac{1}{e^{2\pi i\alpha'X_{38}}-1}\frac{1}{e^{2\pi i\alpha'X_{47}}-1},
    \end{aligned}
\end{equation}

weighted by a factor $\frac{2}{6}$. 

Proceeding now to the computation of the intersection numbers of the second class of two dimensional Stokes polytopes, namely the pentagons, we first note in passing that there are eight squares in total, the intersection numbers of which are obtained by permutations of the indices in the previous formula. For pentagons, there is again a single primitive quadrangulation $((14),(47))$, the $Q$ compatible quadrangulations of which denote the vertices of the Stokes polytope. Listing them we have,

\begin{equation}\label{eq3.11}
    \lbrace{\lbrace{(14),(47)\rbrace},\lbrace{(38),(47)\rbrace},\lbrace{(14),(16)\rbrace},\lbrace{(16),(36)\rbrace},\lbrace{(36),(38)\rbrace}\rbrace}.
\end{equation}

One can confirm readily that this primitive case gives rise to all other pentagons simply by cyclic permutations of the indices. Indeed, the example given in \cite{ref7} is related to our example by moving the indices of our example three steps back, as can be easily checked.

\begin{figure}[h]
    \centering
    \includegraphics[width=0.5\textwidth]{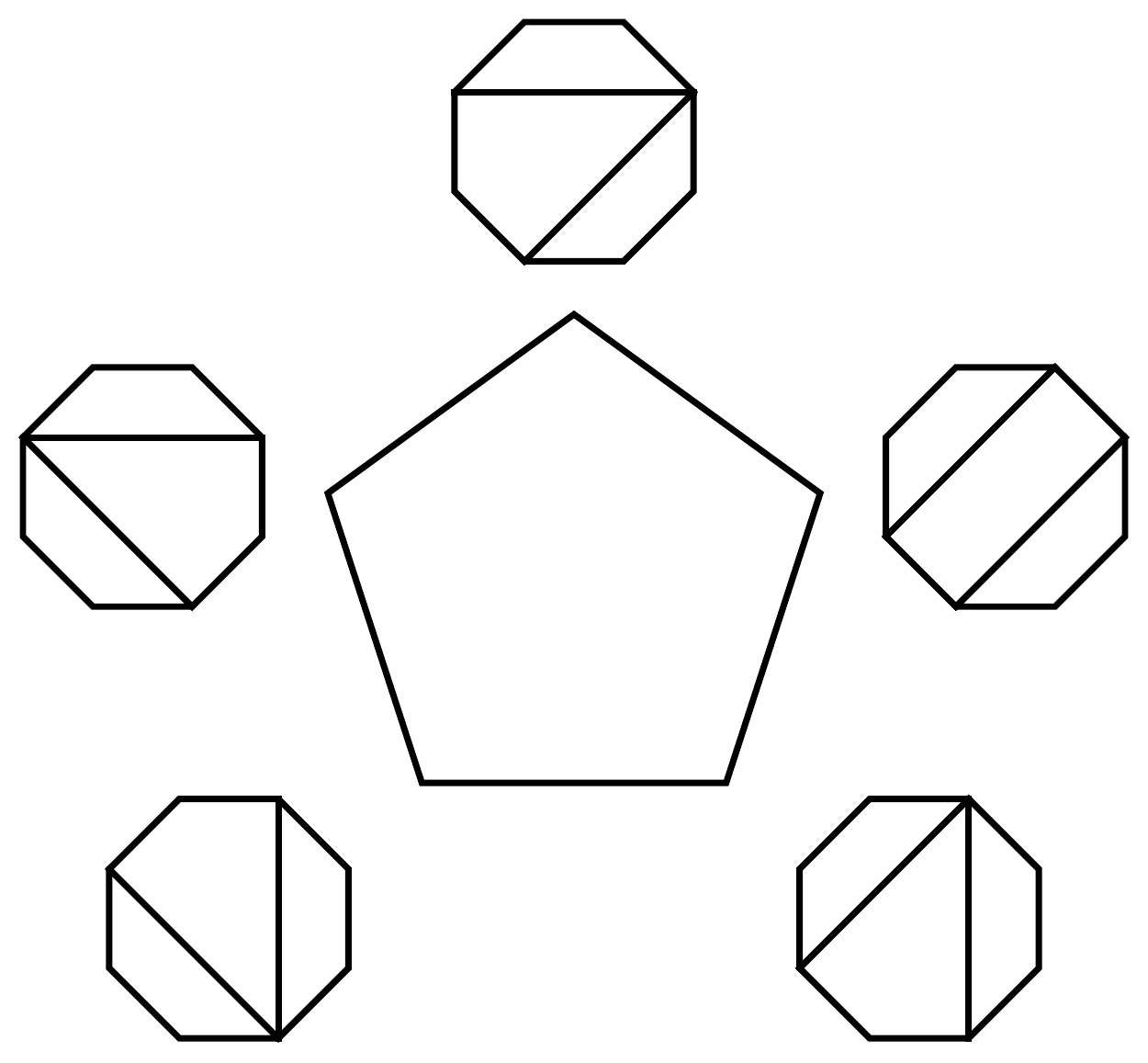}
    \caption{The Stokes polytope given by equation \ref{eq3.11}. The apex vertex is the reference quadrangulation.}
    \label{fig:mesh1}
\end{figure}

Now, the computation of the intersection number can be done in a manner that is identical to the method using which the intersection numbers in the case of the squares were computed. We have contributions from the barycentre, five edges and five vertices for the case of the pentagon, yielding the following,

\begin{equation}
    \begin{aligned}
    & 1 + \frac{1}{e^{2\pi i\alpha' X_{14}} - 1}  + \frac{1}{e^{2\pi i \alpha'X_{47}} - 1} + \frac{1}{e^{2\pi i\alpha' X_{38}} - 1}+ \frac{1}{e^{2\pi i \alpha'X_{16}} - 1}\\& + \frac{1}{e^{2\pi i \alpha'X_{36}} - 1}
    + \frac{1}{e^{2\pi i \alpha'X_{14}} - 1}\frac{1}{e^{2\pi i \alpha'X_{47}} - 1} + \frac{1}{e^{2\pi i \alpha'X_{38}} - 1}\frac{1}{e^{2\pi i \alpha'X_{47}} - 1}\\& + \frac{1}{e^{2\pi i \alpha'X_{14}} - 1}\frac{1}{e^{2\pi i \alpha'X_{16}} - 1}
    + \frac{1}{e^{2\pi i \alpha'X_{16}} - 1}\frac{1}{e^{2\pi i \alpha'X_{36}} - 1}\\& + \frac{1}{e^{2\pi i \alpha'X_{36}} - 1}\frac{1}{e^{2\pi i \alpha'X_{38}} - 1}.
    \end{aligned}
\end{equation}

There are eight such contributions, coming from the eight cyclic permutations that can be obtained from the above intersection number. All of these contributions are now weighted by $\frac{1}{6}$. With this information, we can now write down the full intersection number at this order. The constant term as mentioned already is just a sum over all the weights, which comes out to be $\frac{16}{6}$, from four squares and eight pentagons. Now, the single propagator terms would be obtained by fusing a six point vertex with a four point vertex using a propagator. Correspondingly, we expect that they appear with a factor $\frac{3}{2}$. indeed, this is what arises, since for a given $X_{ij}$, the corresponding single propagator terms arise twice in the squares and five times in the pentagons, giving a total weight factor of $\frac{4}{6}+ \frac{5}{8} = \frac{3}{2}$. Finally, the double propagator terms will all appear with unit residue, due to the definition of the weights. The full details of the computation fot the single propagator case are given in the appendix.

Finally, we note some properties of the intersection numbers in the $10$ point case. Using information regarding the Stokes polytopes for $10$ particle scattering \cite{ref8}\footnote{We are indebted to Prashanth Raman for sharing this information with us.}, it was seen that the constant term in the $10$ particle case came out to be $\frac{125}{24}$. A check was performed by looking at the single propagator terms coming from the fusing of an $8$ particle vertex and a $4$ particle vertex. It was expected that the coefficient be $\frac{16}{6}\lambda^4$, in accordance with the respective coefficients on the vertices. Four different classes of Stokes polytopes arise in the $10$ particle case, so the calculation was tedious, but it was indeed observed that the coefficient matched the expected value.

\section{A Lagrangian Description of the Intersection Theory}
This section will be devoted to providing a Lagrangian description of the theory that is defined by amplitudes computed as intersection numbers of Stokes polytopes. It was seen in the preceding section that the primary building blocks are given by $2n$ point vertices, where $n = 2,3,...$. At each $n$, the vertex comes with a coupling constant given by $\lambda^{n-1}$, where $\lambda$ is the coupling constant in the fiducial $\phi^4$ theory that is described by Stokes polytopes in the field theory limit. In addition to this, one receives a dressing of this vertex factor with a sum over all the weights of the Stokes polytope at that order. Denoting this sum by $\alpha^{\mathcal{S}}_{n}$, we may list the first few as examples,

\begin{equation}
    \alpha^{\mathcal{S}}_{1,2,3} = \frac{3}{2}, \frac{16}{6}, \frac{125}{24}.
\end{equation}

With this notation, the interacting part of the Lagrangian is,

\begin{equation}
   \mathcal{L}^{eff}_{I} = \sum_{n=2}^{\infty}\alpha^{\mathcal{S}}_{n}\lambda^{n-1}\varphi^{2n}.
\end{equation}

The fact that all amplitudes may be derived using the vertices obtained from the above Lagrangian was explicitly observed during the computations of intersection numbers for $6$, $8$ and $10$ particle Stokes polytopes. This assertion admits of a proof however, which makes use of the proof of factorization given in \cite{ref7}. let us recap this for the readers' convenience.

The proof of factorization goes as follows. If we consider a scattering amplitude involving $2n + 4$ particles in $\phi^4$ theory, this is described by a positive geometry of $n$ dimensional Stokes polytopes. Suppose now that a particular channel $X_{ij}$ is allowed to go on the mass shell. This means that the only contribution must come from those Stokes polytopes that contain the diagonal $(ij)$ as a partial quadrangulation. Namely, this means that we have to look at those Stokes polytopes which have facets labelled by this partial quadrangulation. The proof of factorization would now follow if it was established that these facets \emph{each} admit a realization in terms of two lower point Stokes polytopes, which would be given by finding the $Q$ compatible quadrangulations corresponding to dissections of the two lower point polygons obtained as a consequence of singling out the diagonal $(ij)$. The proof of this in \cite{ref7} was supplied by considering a particular convex realization of stokes polytopes, but this happens to be true as a combinatorial fact.

This becomes useful in the following way. The combinatorial and geometric factorization of Stokes polytopes implies in the positive geometry program factorization at the level of amplitudes itself. Unitarity arises as a feature derived from the geometric structure of Stokes polytopes. Even on the mass shell however, we would have to consider the sum over all Stokes polytopes whose reference quadrangulations contain $(ij)$. Thus, in the limit of $X_{ij}\rightarrow 0$ the amplitude contributes a sum over all $\alpha_{Q}$ where $Q$ is a quadrangulation containing $(ij)$ and the factorized amplitude would contribute a product of terms, which would be sums over $\alpha_{Q_1}$ and $\alpha_{Q_2}$, which are the weights from the Stokes polytopes coming from the two lower point polygons obtained by resolving $Q$ into the parts to the left and right of $(ij)$. Thus,

\begin{equation}
    \sum_{(ij)\in Q}\alpha_Q = \sum_{Q_1,Q_2}\alpha_{Q_1}\alpha_{Q_2}.
\end{equation}

The reader will now recognize this as simply a statement of the fact that the intersection number contributions involving propagator factors come from gluing together vertices that involve such sums over $\alpha$'s, so it is sufficient to compute these at all order, which may then be used to compute the intersection numbers.

It is worth recording now the fact that computing these weights is a fairly nontrivial task and has been done effectively only for the cases listed herein. No closed form solution exists so far, although the relation between a possible closed form solution and the above Lagrangian may be interesting to pursue. In \cite{ref1}, a field redefinition was performed to put the Lagrangian for intersecting associahedra in a more compact form, which turns out to be the generating function for Catalan numbers. Catalan numbers are generalized to Fuss-Catalan numbers in the case of quadrangulations. The possibility of finding a point of contact between these two facts will hopefully be realized during future investigation.

\section{Convex Realizations of the Stokes Polytopes}
In this section, we describe how these Stokes polytopes can be described as convex polytopes embedded in complex projective space. In the discussion to follow, we will restrict ourselves to the real parts in order to facilitate ease of visualization, but it is hoped that no confusion will arise as a result.

Defining the convex description requires the introduction of some formalism, the development of which will follow \cite{ref9} closely. Let us first recall that a given Feynman diagram in $\phi^4$ theory is associated to a quadrangulation of an $(2n+4)$-gon. Let us call this dissection $Q$. Now, define a dual $(2n+4)$-gon whose vertices are obtained by rotating the initial $(2n+4)$-gon by $\frac{\pi}{2n+4}$ clockwise\footnote{Visualizing the vertex $i$ as located at the center of edge $i,i+1$ might be easier for the reader. }. This is normally called the solid polygon and the original would be called the hollow polygon. Denoting now the reference quadrangulation by $Q$, we will call all the corresponding $Q$-compatible quadrangulations by $Q_1,\dots,Q_{k}$. Let $\delta_1,\dots,\delta_{p}$ be the set of all diagonals that appear in these quadrangulations. 

Recall that the diagonals defined combinatorially the facets of the Stokes polytope. Thus, given a diagonal and the reference quadrangulation, there exists a quantity that tells us how the corresponding facet manifests as an embedding, and this quantity is called the $d$ vector. The $d$ vector is defined in $\mathbb{R}^{n}$, where the basis elements are labelled by the diagonals $\overline{\delta}_1,\dots,\overline{\delta}_n$ making up the reference dissection. Now the $d$ vector is defined as follows. Given the reference quadrangulation $Q$ of the hollow polygon and a dissection $\delta$ on the solid polygon and let $\overline{\delta}$ be a diagonal in $Q$ crossed by $\delta$. Note now that $Q$ is split into two parts by $\overline{\delta}$. Including now the edges of $Q$ in addition to the diagonals, let $\mu$ and $\nu$ be the two edges or diagonals crossed by $\delta$ on the two cells of $Q$ defined by $\overline{\delta}$. If $\mu\overline{\delta}\nu$ is a path, then define $\varepsilon(\delta|\overline{\delta})$ as follows. If the path resembles $Z$, then it is $-1$, if it resembles a $Z$ laterally inverted it is $1$ and if it resembles $\VVV$ then it is zero.

The $d$ vector of $\delta$ is now,

\begin{equation}
    d_{\delta} = \sum_{\overline{\delta}\in Q}\varepsilon(\delta|\overline{\delta})\hat{e}_{\overline{\delta}},
\end{equation}

where $\hat{e}_{\overline{\delta}}$ is the basis element corresponding to the diagonal $\overline{\delta}$.

Although this would seem rather formal to the reader, a simple exercise is instructive. Consider the reference quadrangulation corresponding to $(14)$ for the six particle case. There were two diagonals now that gave rise to quadrangulations compatible with this, namely $(14)$ and $(36)$. A simple calculation will show that for $(14)$, the $d$ vector is $-\hat{e}_{(14)}$ and for $(36)$ the $d$ vector is $\hat{e}_{(14)}$.

Now, denoting the number of diagonals in $Q$ crossed by $\delta$ as $w(\delta|Q)$, the facets of the Stokes polytopes are given by,

\begin{equation}
    x\cdot d_{\delta} \leq w(\delta|Q),
\end{equation}

where $x = x_1\hat{e}_{\overline{\delta}_1}+\dots+x_n\hat{e}_{\overline{\delta}_n}$. Thus for the case at hand, namely the Stokes polytope for $(14)$, we get,

\begin{equation}
    -x \leq 1
\end{equation}

and 

\begin{equation}
    x \leq 1,
\end{equation}

which indeed is the line segment from $-1$ to $1$, which is the one dimensional Stokes polytope. This fact inspired the unusual choice made earlier of removing points $-1,1,\infty$ during the calculations.

\section{The Field Theory Limit}\label{sec6}
\subsection{The Field Theory Limit of Intersection Numbers}
We are now in the ideal position to explore the limit of vanishing $\alpha'$ in the context of the intersection numbers computed in previous sections. It should be emphasized that this $\alpha'$, though seemingly suggestive, was introduced as a regulator, in order to ensure that the exponentials carried dimensionless quantities. Nevertheless, the limit in which this vanishes is interesting, as field theory amplitudes are recovered, as we shall now see.

In general, supposing we have a scattering process involving $2n+4$ particles, for $n$ starting from $0$, the Stokes polytopes needed to compute the amplitudes will be $n$ dimensional. Now, if we were to compute the intersection numbers, we will obtain contributions from boundaries of all codimensions, involving $k$ propagator terms for codimension $k$. Recall first that a propagator is of the form,

\begin{equation}
    \frac{1}{e^{i\alpha'X_{ij}}-1}.
\end{equation}

In the $\alpha'\rightarrow 0$ limit, it is clear that these propagators obey the singular behaviour,

\begin{equation}
    \frac{1}{i\alpha' X_{ij}}.
\end{equation}

Consequently, the most singular contribution to intersection numbers will come from those terms arising from the boundaries of highest codimension. Indeed, these are the vertices of the Stokes polytopes, which encode the maximal quadrangulations of the $2n+4$ gon, and when summed over by weighting them appropriately, the full $\phi^4$ amplitude in the planar limit will be obtained. Thus, intersection numbers of Stokes polytopes, which we denote $\langle{\mathrm{Stokes}_{n,Q},\mathrm{Stokes}_{n,Q}}\rangle$ yield in the vanishing $\alpha'$ limit,

\begin{equation}
    \sum_{Q}\alpha_Q \langle{\mathrm{Stokes}_{n,Q},\mathrm{Stokes}_{n,Q}}\rangle =_{\alpha'\rightarrow 0} \frac{1}{(\alpha')^n}m^{\phi^4}_{2n+4}.
\end{equation}

To explain the notation, we have simply summed over the intersection numbers of all Stokes polytopes of dimension $n$ by weighting with the coefficients $\alpha_{Q}$, yielding the planar field theory amplitude for $\phi^4$ theory for the scattering of $2n+4$ particles.

Alternatively, the field theory limit, which is a term we can now responsibly use to refer to the $\alpha'\rightarrow 0$ limit, can be attained by studying the intersection theory of twisted cocycles rather that than that of twisted cycles. To do this, we employ the formalism laid out in \cite{ref1-1}. We will not ponder the mathematical details, but in a nutshell, while the intersection theory of twisted cocycles explores the full configuration space of the hyperplane arrangement, the cohomology classes of the twist differential operator admit an intersection theory as well, which only probes contributions from the boundaries. In order to illustrate this, we will look at a few examples.

Let us first compute the intersection numbers of cycles corresponding to the one dimensional Stokes polytope. We will assume in our subsequent calculations that the reader is familiar with the procedure delineated in \cite{ref1-1}. We define the following differential form on $X = \mathbb{CP}^{1}-\lbrace{1,-1,\infty\rbrace}$\footnote{The reader may note that the coefficient of the point at infinity would be $-X_{14} - X_{36}$, commensurate with the discussion had earlier regarding this point.}. For the Stokes polytope corresponding to $(14)$, we look at the following twist,

\begin{equation}
    \omega = d\log(z+1)^{X_{14}} + d\log(1-z)^{X_{36}} = \omega_{z}dz,
\end{equation}

where,

\begin{equation}
    \omega_{z} = \frac{X_{14}}{1+z}+\frac{X_{36}}{1-z} .
\end{equation}

Defining now the one form,

\begin{equation}
    \varphi_{(14),x}dz = d\log\left(\frac{1+z}{1-z}\right)
\end{equation}

the twisted intersection number defined as,

\begin{equation}
    \int_{X}\delta(\omega_{z})\varphi_{(14),z}\varphi_{(14),z}dz
\end{equation}

yields,

\begin{equation}
    \frac{1}{X_{14}} + \frac{1}{X_{36}},
\end{equation}

upto a global sign. Now, a sum over all the Stokes polytopes corresponding to all three quadrangulations weighted appropriately will give the amplitude for six particle scattering.

Now for eight particle scattering, we will look at the square Stokes polytope, as the pentagon case will follow naturally. For the square corresponding to quadrangulation $(14,58)$, we require a hyperplane arrangement composed of four hyperplanes as indicated in section \ref{sec3.2}. The twist is,

\begin{equation}
\begin{aligned}
    &X_{14}d\log(H_{(14)}) + X_{58}d\log(H_{(58)}) + X_{38}d\log(H_{(38)}) + X_{47}d\log(H_{(47)})\\& = \omega_{z_1}dz_1 + \omega_{z_2}dz_2,
\end{aligned}
\end{equation}

where $z_1,z_2$ are inhomogeneous coordinates on $X = \mathbb{CP}^2 - \lbrace{H_{(14)},H_{(58)},H_{(38)},H_{(47)}}\rbrace - \lbrace{H_{\infty 1},H_{\infty 2}\rbrace}$, where we have indicated removing hyperplanes pushed to infinity as well. Now the twisted form to be used is,

\begin{equation}
\begin{aligned}
\varphi_{(14,58)}& = d\log\left(\frac{H_{(14)}}{H_{(58)}}\right)\wedge d\log\left(\frac{H_{(58)}}{H_{(47)}}\right) - d\log\left(\frac{H_{(38)}}{H_{(58)}}\right)\wedge d\log\left(\frac{H_{(58)}}{H_{(47)}}\right)\\ &= \hat{\varphi}_{(14,58)}dz_1\wedge dz_2,
\end{aligned}
\end{equation}

where we have explicitly projectivized the form\footnote{The reader may notice that the forms must necessarily come from an overcomplete basis of the twisted cohomology group, a trade-off due to the fact that Stokes polytopes are not unique.}. The intersection number is now,

\begin{equation}
    \int_{X} \delta(\omega_{z_1})\delta(\omega_{z_2})\hat{\varphi}_{(14,58)}\hat{\varphi}_{(14,58)}dz_1dz_2.
\end{equation}

Summing over all such intersection numbers from all polytopes weighted correctly will yield the amplitude for eight particle scattering, as the reader may readily check.

\subsection{Quartic Vertices in String Theory}
In this short section, we will comment on the apparent inability of string theory amplitudes to deal properly with quartic vertices and will point out how our approach in this paper represents a small improvement on the state of affairs.

The attempt to study the presence of quartic vertices in string theory is not a recent one. In particular, a curious approach was taken in \cite{ref10}, which we will now highlight and contrast with the route taken in the present work.

Let us focus on the six particle case considered by the authors of \cite{ref10}. To describe the scattering of spinless particles in open bosonic string theory, the disk integrals of the tachyonic sector have to be considered. The moduli space is this case is the disk marked with $6$ points, thus delineating three parameters of integration $z_2,z_3,z_4$, with $z_1=0$, $z_5=1$ and $z_6$ pushed to infinity. 

How does one now recover a quartic vertex? It is expected that any field theory contribution from a string integral must come from isolated points in the moduli space. Accordingly, one would expect that the quartic interaction arising from the interaction of particles $1,2,3$, namely the $X_{14}$ channel would come from that region of the moduli space probed when $z_2,z_3$ approach the neighbourhood of $z_1$ and $z_5$ approaches $z_5$. It was shown in the aforementioned reference by the authors that this region of the moduli space is directly accessed by effecting the following transformation in the field theory limit,

\begin{equation}
    \begin{aligned}
    z_2 &= e^{-\frac{t}{\alpha'}},\\
    z_3 &= ye^{-\frac{t}{\alpha'}},\\
    z_4 &= x.
    \end{aligned}
\end{equation}

In this limit, it can be established by direct computation that the terms unsuppressed by $\alpha'$ yield up to a colour ordering the amplitude,

\begin{equation}
    \sim \frac{1}{(p_1+p_2+p_3)^2 - m^2},
\end{equation}

where the mass is tachyonic.

It is instructive to compare this computational approach with the one advocated in this work. There are two points which may be raised. On the one hand, Stokes polytopes are attractive as they seem to furnish an \emph{a priori} moduli space for pure quartic interactions. In the string theory case, the regions of integration encoding the quartic interactions are subspaces of the full moduli space and have to be probed in that light. On the other hand, the string theory approach lends itself to ready generalization to the case of higher loop interactions. Analogous regions contained in $\mathcal{M}_{g,n}$ encode information about higher loop interactions with quartic vertices. No such picture has yet been uncovered for the case of pure quartic interactions, \emph{viz}, a higher 'genus' analogue of Stokes polytopes. Consequently, this may serve as an interesting route to further investigation. It is gratifying however that at least at tree level, there does exist a moduli space for quartic interactions independent of traditional string theory moduli space constructions, which admits an $\alpha'$ deformation.

\section{Massive Amplitudes}
So far, our analysis has been restricted to the case under which the particles involved in the scattering processes are massive. Although the inclusion of mass has often presented a difficulty in the context of CHY and intersection theory amplitudes, some progress has been reported (see \cite{mizerathesis}). In this work however, we believe that the formalism presented here is particularly conducive to the inclusion of masses, and this is traced to the nature of the convex realization of the polytopes involved.

Traditionally, at least in the case of CHY, the associahedron is realized as the Deligne-Mumford compactification of the simplex, which is embedded in $\mathbb{CP}^{n}$ through a hyperplane arrangement. In our approach however, each factorization channel involved in the planar amplitude is associated to a given facet of the polytope, which finds its convex realization as a hyperplane in $\mathbb{CP}^{n}$, to which we assign a mass by simply shifting the coefficient from $X_{ij}$ to $X_{ij} + m^{2}_{ij}$. In this manner, we do not have to worry about the residue at infinity or any contraints therefrom, as these can simply be removed by an addition of hyperplanes at infinity.

Let us illustrate this with the case of $8$ particle scattering in $\phi^4$ theory. In order to make the treatment as concrete as possible, we explicitly give the hyperplane arrangement for the case of the square Stokes polytope Considered in section III B.

The square Stokes polytope in $2$ dimensional, requiring hence the ambient space $\mathbb{CP}^{2}$. Now the reference dissection is $(14,58)$. In order to find the convex realization of the Stokes polytope, we have to compute the $d$ vectors of the facets $(14)$, $(58)$, $(38)$ and $(47)$. 

Let us begin with $(14)$. The figure required is as follows.

\begin{figure}[h]
    \centering
    \includegraphics[width=0.5\textwidth]{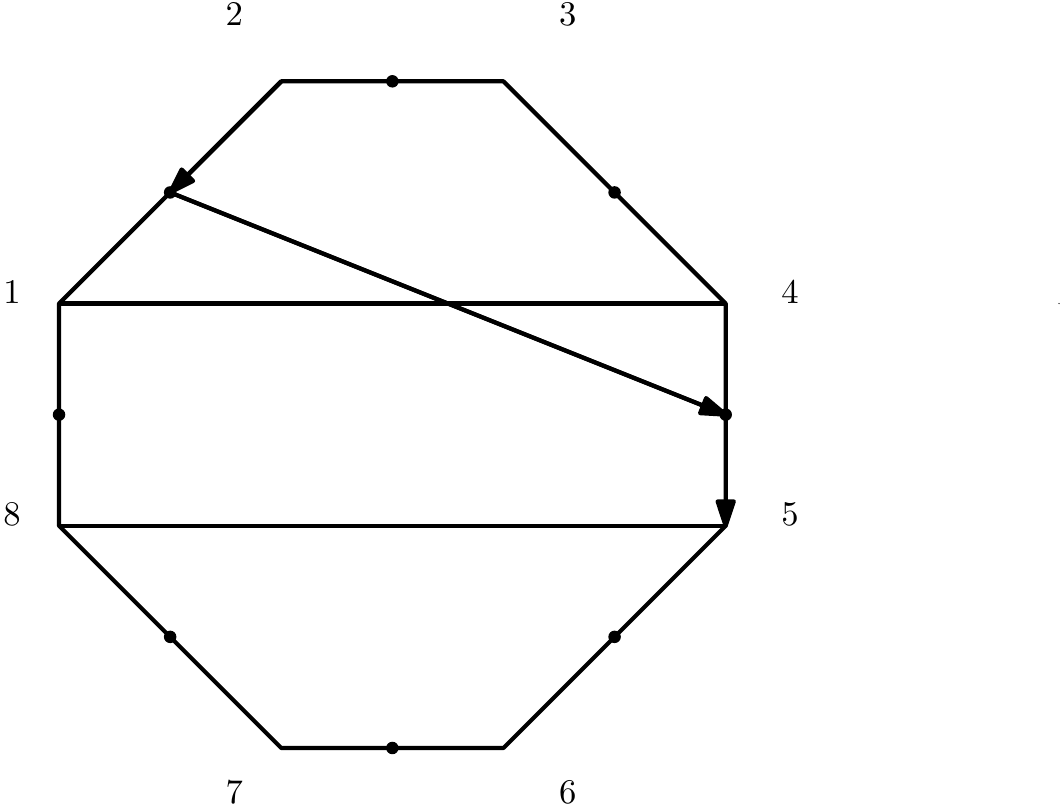}
    \caption{The arrow indicates the direction of the path to be considered.}
    \label{fig:mesh1}
\end{figure}

In the figure, the reference dissection is $(14,58)$, with the solid dissection $(1'4')$ indicated by the internal arrow cutting $(14)$. Now, the path is a laterally inverted $Z$ with one reference diagonal being cut. Hence, the $d$ vector is $\hat{e}_{(14)}$ and the facet corresponding to this channel is,

\begin{equation}
    (x_{1}\hat{e}_{(14)} + x_2\hat{e}_{(58)})\cdot \hat{e}_{(14)} \leq 1,
\end{equation}

which gives,

\begin{equation}
    x_1 \leq 1.
\end{equation}

The other facets corresponding to $(58)$, $(38)$ and $(47)$ are obtained in the same manner, and are:

\begin{equation}
    x_2 \leq 1,
\end{equation}

\begin{equation}
    x_{1} \geq -1
\end{equation}
and
\begin{equation}
    x_{2} \geq -1,
\end{equation}

clearly bounding a square.

\vspace{0.3cm}

\begin{figure}[h]
    \centering
    \includegraphics[width=0.4\textwidth]{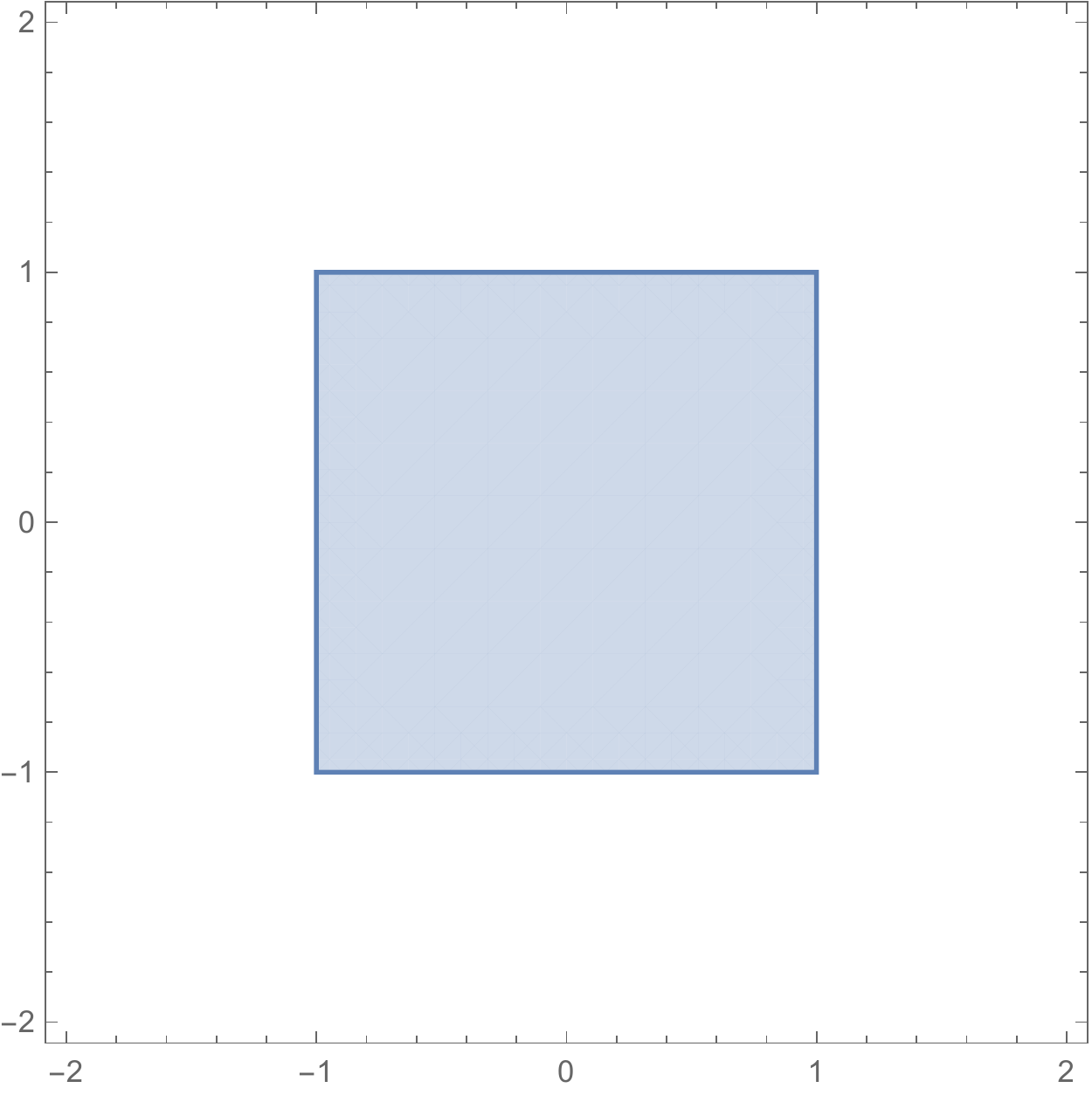}
    \caption{The Stokes polytope corresponding to $(14,58)$}
    \label{fig:mesh1}
\end{figure}

Now, working in inhomogeneous coordinates $(x,y)$ of $\mathbb{CP}^{n}$, we can use the standard tools of twisted intersection theory on,

\begin{equation}
    X = \mathbb{CP}^{n} - \lbrace{f_{1} = x-1 = 0, f_{2} = y-1 = 0, f_{3} = x+1=0, f_{4} = y+1=0,H_{\infty,x},H_{\infty,y}\rbrace}.
\end{equation}

Two hyperplanes are removed at infinity in order to suitably cancel residues as we shall soon see. With this configuration space, we now define the twist,

\begin{equation}
    \omega = (X_{14}-m^{2}_{14})d\log f_{1} +  (X_{58}-m^{2}_{58})d\log f_{2} +  (X_{38}-m^{2}_{38})d\log f_{3}+  (X_{47}-m^{2}_{47})d\log f_{4}.
\end{equation}

In order to remove the residues at infinity, the weight carried by the hyperplane at $x \rightarrow \infty$ is chosen to be $-X_{14}-X_{38}+m^{2}_{14}+m^{2}_{38}$ and that carried out by the hyperlane at $y\rightarrow \infty$ is $-X_{58}-X_{47}+m^{2}_{58}+m^{2}_{47}$.

To compute the contribution to the amplitude from this Stokes polytope, we choose the following element of the twisted cohomology,

\begin{equation}
    \phi_{(14,58)} = d\log{\frac{f_1}{f_2}}\wedge d\log{\frac{f_2}{f_4}} - d\log{\frac{f_3}{f_2}}\wedge d\log{\frac{f_2}{f_4}}.
\end{equation}

The self intersection number of this is computed by formula $(38)$, but now instead of the massless amplitude, we obtain the amplitude,

\begin{equation}
\begin{aligned}
     &\frac{1}{X_{14}-m^2_{14}}\frac{1}{X_{58}-m^2_{58}} + \frac{1}{X_{58}-m^2_{58}}\frac{1}{X_{38}-m^2_{38}} +\\ &\frac{1}{X_{38}-m^2_{38}}\frac{1}{X_{47}-m^2_{47}} + 
    \frac{1}{X_{47}-m^2_{47}}\frac{1}{X_{14}-m^2_{14}}.
\end{aligned}
\end{equation}

It should be clear from this discussion that the formalism presented in this section is readily extended to any amplitude that can be computed using the technology of Stokes polytopes, the only obstruction being the practical construction of the Stokes polytopes being considered.
\section{Discussion}
Investigating the relationship between worldsheet approaches to quantum field theory and higher point vertices has certainly proved to be a fruitful area of investigation. Stokes polytopes have been extremely useful in their ability to help us probe the worldsheet structure of quartic interactions in a manner that is both expedient and capable of making contact with mathematically rigorous approaches to worldsheet integrals, namely intersection theory. The latter point has been our main focus in this article. As we have seen, quartic interactions of scalar particles admit a string-like deformation in a manner reminiscent of the relation borne by the inverse string KLT kernel to biadjoint amplitudes.

There are several clear directions in which we may proceed to understand this line of investigation more thoroughly, which we will now lay out. An investigation in progress \cite{alok2} is in an effort to understand how the canonical forms of Stokes polytopes descend from the embeddings of these objects in associahedra. There is evidence to suggest that associahedra are in a precise sense more fundamental, and that the embeddings of these objects in kinematical space determine uniquely how Stokes polytopes are embedded therein. Bringing into agreement the convex structure of Stokes polytopes presented as hyperplane arrangements in the present work with the understanding gleaned from viewing these as lying inside associahedra is one line of work that will definitely be interesting to address.

Indeed, this suggests that a realization of geometric structures encoding data about quartic and higher point vertices at higher loops may well be within sight. If the understanding of string theory at the worldsheet for genus zero is indeed sufficient to detail quartic and higher point interactions at tree level, exploring the higher genus analogues might be quite revealing. At the moment, the one loop structure of cubic diagrams in the positive geometry program \cite{halo} has been uncovered. Hopefully, investigations along the general directions just highlighted may prove fruitful in going beyond this.

Finally, in a direction somewhat orthogonal to the ones discussed so far, applying the chain of reasoning presented in this article to the accordiohedra discussed in \cite{nikhil} is an obvious step to be taken. Indeed, the twisted intersection theory of cocycles associated to accordiohedra would be expected to supply tree level amplitudes for generic scalar theories, so explicitly working this out is a problem that is of interest.

\section*{Acknowledgements}

It is a pleasure to acknowledge gratitude to Alok Laddha for his careful reading of the manuscript. The author would also like to thank Oliver Schlotterer for his invaluable comments and Sudip Ghosh for interesting discussions. The author thanks Freddy Cachazo, Wan Zhen Chua, Nick Early, Alfredo Guevara, Sebastian Mizera, Seyed Faroogh  Moosavian,  Yasha  Neiman, Prashanth Raman and Vasudev  Shyam, with special thanks to Wan Zhen Chua for help with the diagrams, to Yasha Neiman and the Quantum Gravity Group at OIST for support during the early stages of work and especially to Prashanth Raman, for his generous supply of data related to Stokes Polytopes.

This work was supported by a scholarship made possible by the Perimeter Scholars International Program. Research at the Perimeter Institute is funded by the Government of Canada via Industry Canada as well as by the Province of Ontario via the Ministry of Economic Development and Innovation.

\appendix

\section{Appendix}

In this appendix, we record all the intersection numbers to be computed in the $n=8$ case, demonstrating in the process how the coefficients of the final result come about. 

Recapping first the intersection number of the Stokes polytope for the quadrangulation $(14,58)$,

\begin{equation}
    \begin{aligned}
    &\langle{\mathcal{S}_{2,(14,58)}\otimes\mathrm{Stokes}_{2, (14,58)}(z_1,z_2),\mathcal{S}_{2,(14,58)}\otimes\mathrm{Stokes}_{2, (14,58)}(z_1,z_2)\rangle}=\\
       & 1 + \frac{1}{e^{2\pi i\alpha'X_{14}}-1} + \frac{1}{e^{2\pi i\alpha'X_{38}}-1}  + \frac{1}{e^{2\pi i\alpha'X_{47}}-1} + \frac{1}{e^{2\pi i\alpha'X_{58}}-1}\\
       &+ \frac{1}{e^{2\pi i\alpha'X_{14}}-1}\frac{1}{e^{2\pi i\alpha'X_{58}}-1}
         + \frac{1}{e^{2\pi i\alpha'X_{14}}-1}\frac{1}{e^{2\pi i\alpha'X_{47}}-1}\\& + \frac{1}{e^{2\pi i\alpha'X_{38}}-1}\frac{1}{e^{2\pi i\alpha'X_{58}}-1}
         + \frac{1}{e^{2\pi i\alpha'X_{38}}-1}\frac{1}{e^{2\pi i\alpha'X_{47}}-1},
    \end{aligned}
\end{equation}

which can now be used to generate all the intersection numbers for Stokes polytopes arising from quadrangulations of this type, namely $(25,16)$, $(36,27)$ and $(47,38)$. Explicitly they read,

\begin{equation}
    \begin{aligned}
    &\langle{\mathcal{S}_{2,(25,16)}\otimes\mathrm{Stokes}_{2, (25,16)}(z_1,z_2),\mathcal{S}_{2,(25,16)}\otimes\mathrm{Stokes}_{2, (25,16)}(z_1,z_2)\rangle}=\\
       & 1 + \frac{1}{e^{2\pi i\alpha'X_{25}}-1} + \frac{1}{e^{2\pi i\alpha'X_{14}}-1}  + \frac{1}{e^{2\pi i\alpha'X_{58}}-1} + \frac{1}{e^{2\pi i\alpha'X_{16}}-1}\\
       &+ \frac{1}{e^{2\pi i\alpha'X_{25}}-1}\frac{1}{e^{2\pi i\alpha'X_{16}}-1}
         + \frac{1}{e^{2\pi i\alpha'X_{25}}-1}\frac{1}{e^{2\pi i\alpha'X_{58}}-1}\\& + \frac{1}{e^{2\pi i\alpha'X_{14}}-1}\frac{1}{e^{2\pi i\alpha'X_{16}}-1}
         + \frac{1}{e^{2\pi i\alpha'X_{14}}-1}\frac{1}{e^{2\pi i\alpha'X_{58}}-1},
    \end{aligned}
\end{equation}

\begin{equation}
    \begin{aligned}
    &\langle{\mathcal{S}_{2,(36,27)}\otimes\mathrm{Stokes}_{2, (36,27)}(z_1,z_2),\mathcal{S}_{2,(36,27)}\otimes\mathrm{Stokes}_{2, (36,27)}(z_1,z_2)\rangle}=\\
       & 1 + \frac{1}{e^{2\pi i\alpha'X_{36}}-1} + \frac{1}{e^{2\pi i\alpha'X_{25}}-1}  + \frac{1}{e^{2\pi i\alpha'X_{16}}-1} + \frac{1}{e^{2\pi i\alpha'X_{27}}-1}\\
       &+ \frac{1}{e^{2\pi i\alpha'X_{36}}-1}\frac{1}{e^{2\pi i\alpha'X_{27}}-1}
         + \frac{1}{e^{2\pi i\alpha'X_{36}}-1}\frac{1}{e^{2\pi i\alpha'X_{16}}-1}\\& + \frac{1}{e^{2\pi i\alpha'X_{25}}-1}\frac{1}{e^{2\pi i\alpha'X_{27}}-1}
         + \frac{1}{e^{2\pi i\alpha'X_{25}}-1}\frac{1}{e^{2\pi i\alpha'X_{16}}-1},
    \end{aligned}
\end{equation}

and

\begin{equation}
    \begin{aligned}
    &\langle{\mathcal{S}_{2,(47,38)}\otimes\mathrm{Stokes}_{2, (47,38)}(z_1,z_2),\mathcal{S}_{2,(47,38)}\otimes\mathrm{Stokes}_{2, (47,38)}(z_1,z_2)\rangle}=\\
       & 1 + \frac{1}{e^{2\pi i\alpha'X_{47}}-1} + \frac{1}{e^{2\pi i\alpha'X_{36}}-1}  + \frac{1}{e^{2\pi i\alpha'X_{27}}-1} + \frac{1}{e^{2\pi i\alpha'X_{38}}-1}\\
       &+ \frac{1}{e^{2\pi i\alpha'X_{47}}-1}\frac{1}{e^{2\pi i\alpha'X_{38}}-1}
         + \frac{1}{e^{2\pi i\alpha'X_{47}}-1}\frac{1}{e^{2\pi i\alpha'X_{27}}-1}\\& + \frac{1}{e^{2\pi i\alpha'X_{36}}-1}\frac{1}{e^{2\pi i\alpha'X_{38}}-1}
         + \frac{1}{e^{2\pi i\alpha'X_{36}}-1}\frac{1}{e^{2\pi i\alpha'X_{27}}-1}.
    \end{aligned}
\end{equation}

Starting now from the quadrangulation $(14,47)$, we can obtain the intersection numbers for the pentagons. For this quadrangulation we have,

\begin{equation}
    \begin{aligned}
    &\langle{\mathcal{S}_{2,(14,47)}\otimes\mathrm{Stokes}_{2, (14,47)}(z_1,z_2),\mathcal{S}_{2,(14,47)}\otimes\mathrm{Stokes}_{2, (14,47)}(z_1,z_2)\rangle}=\\
    & 1 + \frac{1}{e^{2\pi i\alpha' X_{14}} - 1}  + \frac{1}{e^{2\pi i \alpha'X_{47}} - 1} + \frac{1}{e^{2\pi i\alpha' X_{38}} - 1}+ \frac{1}{e^{2\pi i \alpha'X_{16}} - 1}\\& + \frac{1}{e^{2\pi i \alpha'X_{36}} - 1}
    + \frac{1}{e^{2\pi i \alpha'X_{14}} - 1}\frac{1}{e^{2\pi i \alpha'X_{47}} - 1} + \frac{1}{e^{2\pi i \alpha'X_{38}} - 1}\frac{1}{e^{2\pi i \alpha'X_{47}} - 1}\\& + \frac{1}{e^{2\pi i \alpha'X_{14}} - 1}\frac{1}{e^{2\pi i \alpha'X_{16}} - 1}
    + \frac{1}{e^{2\pi i \alpha'X_{16}} - 1}\frac{1}{e^{2\pi i \alpha'X_{36}} - 1}\\& + \frac{1}{e^{2\pi i \alpha'X_{36}} - 1}\frac{1}{e^{2\pi i \alpha'X_{38}} - 1}.
    \end{aligned}
\end{equation}

Permuting indices we obtain the intersection numbers for all pentagons, which we record here,

\begin{equation}
    \begin{aligned}
    &\langle{\mathcal{S}_{2,(25,58)}\otimes\mathrm{Stokes}_{2, (25,58)}(z_1,z_2),\mathcal{S}_{2,(25,58)}\otimes\mathrm{Stokes}_{2, (25,58)}(z_1,z_2)\rangle}=\\
    & 1 + \frac{1}{e^{2\pi i\alpha' X_{25}} - 1}  + \frac{1}{e^{2\pi i \alpha'X_{58}} - 1} + \frac{1}{e^{2\pi i\alpha' X_{14}} - 1}+ \frac{1}{e^{2\pi i \alpha'X_{27}} - 1}\\& + \frac{1}{e^{2\pi i \alpha'X_{47}} - 1}
    + \frac{1}{e^{2\pi i \alpha'X_{25}} - 1}\frac{1}{e^{2\pi i \alpha'X_{58}} - 1} + \frac{1}{e^{2\pi i \alpha'X_{14}} - 1}\frac{1}{e^{2\pi i \alpha'X_{58}} - 1}\\& + \frac{1}{e^{2\pi i \alpha'X_{25}} - 1}\frac{1}{e^{2\pi i \alpha'X_{27}} - 1}
    + \frac{1}{e^{2\pi i \alpha'X_{27}} - 1}\frac{1}{e^{2\pi i \alpha'X_{47}} - 1}\\& + \frac{1}{e^{2\pi i \alpha'X_{47}} - 1}\frac{1}{e^{2\pi i \alpha'X_{14}} - 1},
    \end{aligned}
\end{equation}

\begin{equation}
    \begin{aligned}
    &\langle{\mathcal{S}_{2,(16,36)}\otimes\mathrm{Stokes}_{2, (16,36)}(z_1,z_2),\mathcal{S}_{2,(16,36)}\otimes\mathrm{Stokes}_{2, (16,36)}(z_1,z_2)\rangle}=\\
    & 1 + \frac{1}{e^{2\pi i\alpha' X_{36}} - 1}  + \frac{1}{e^{2\pi i \alpha'X_{16}} - 1} + \frac{1}{e^{2\pi i\alpha' X_{25}} - 1}+ \frac{1}{e^{2\pi i \alpha'X_{38}} - 1}\\& + \frac{1}{e^{2\pi i \alpha'X_{58}} - 1}
    + \frac{1}{e^{2\pi i \alpha'X_{36}} - 1}\frac{1}{e^{2\pi i \alpha'X_{16}} - 1} + \frac{1}{e^{2\pi i \alpha'X_{25}} - 1}\frac{1}{e^{2\pi i \alpha'X_{16}} - 1}\\& + \frac{1}{e^{2\pi i \alpha'X_{36}} - 1}\frac{1}{e^{2\pi i \alpha'X_{38}} - 1}
    + \frac{1}{e^{2\pi i \alpha'X_{38}} - 1}\frac{1}{e^{2\pi i \alpha'X_{58}} - 1}\\& + \frac{1}{e^{2\pi i \alpha'X_{58}} - 1}\frac{1}{e^{2\pi i \alpha'X_{25}} - 1},
    \end{aligned}
\end{equation}

\begin{equation}
    \begin{aligned}
    &\langle{\mathcal{S}_{2,(27,47)}\otimes\mathrm{Stokes}_{2, (27,47)}(z_1,z_2),\mathcal{S}_{2,(27,47)}\otimes\mathrm{Stokes}_{2, (27,47)}(z_1,z_2)\rangle}=\\
    & 1 + \frac{1}{e^{2\pi i\alpha' X_{47}} - 1}  + \frac{1}{e^{2\pi i \alpha'X_{27}} - 1} + \frac{1}{e^{2\pi i\alpha' X_{36}} - 1}+ \frac{1}{e^{2\pi i \alpha'X_{14}} - 1}\\& + \frac{1}{e^{2\pi i \alpha'X_{16}} - 1}
    + \frac{1}{e^{2\pi i \alpha'X_{47}} - 1}\frac{1}{e^{2\pi i \alpha'X_{27}} - 1} + \frac{1}{e^{2\pi i \alpha'X_{36}} - 1}\frac{1}{e^{2\pi i \alpha'X_{27}} - 1}\\& + \frac{1}{e^{2\pi i \alpha'X_{47}} - 1}\frac{1}{e^{2\pi i \alpha'X_{14}} - 1}
    + \frac{1}{e^{2\pi i \alpha'X_{14}} - 1}\frac{1}{e^{2\pi i \alpha'X_{16}} - 1}\\& + \frac{1}{e^{2\pi i \alpha'X_{16}} - 1}\frac{1}{e^{2\pi i \alpha'X_{36}} - 1},
    \end{aligned}
\end{equation}

\begin{equation}
    \begin{aligned}
    &\langle{\mathcal{S}_{2,(38,58)}\otimes\mathrm{Stokes}_{2, (38,58)}(z_1,z_2),\mathcal{S}_{2,(38,58)}\otimes\mathrm{Stokes}_{2, (38,58)}(z_1,z_2)\rangle}=\\
    & 1 + \frac{1}{e^{2\pi i\alpha' X_{58}} - 1}  + \frac{1}{e^{2\pi i \alpha'X_{38}} - 1} + \frac{1}{e^{2\pi i\alpha' X_{47}} - 1}+ \frac{1}{e^{2\pi i \alpha'X_{25}} - 1}\\& + \frac{1}{e^{2\pi i \alpha'X_{27}} - 1}
    + \frac{1}{e^{2\pi i \alpha'X_{58}} - 1}\frac{1}{e^{2\pi i \alpha'X_{38}} - 1} + \frac{1}{e^{2\pi i \alpha'X_{47}} - 1}\frac{1}{e^{2\pi i \alpha'X_{38}} - 1}\\& + \frac{1}{e^{2\pi i \alpha'X_{58}} - 1}\frac{1}{e^{2\pi i \alpha'X_{25}} - 1}
    + \frac{1}{e^{2\pi i \alpha'X_{25}} - 1}\frac{1}{e^{2\pi i \alpha'X_{27}} - 1}\\& + \frac{1}{e^{2\pi i \alpha'X_{27}} - 1}\frac{1}{e^{2\pi i \alpha'X_{47}} - 1},
    \end{aligned}
\end{equation}

\begin{equation}
    \begin{aligned}
    &\langle{\mathcal{S}_{2,(14,16)}\otimes\mathrm{Stokes}_{2, (14,16)}(z_1,z_2),\mathcal{S}_{2,(14,16)}\otimes\mathrm{Stokes}_{2, (14,16)}(z_1,z_2)\rangle}=\\
    & 1 + \frac{1}{e^{2\pi i\alpha' X_{16}} - 1}  + \frac{1}{e^{2\pi i \alpha'X_{14}} - 1} + \frac{1}{e^{2\pi i\alpha' X_{58}} - 1}+ \frac{1}{e^{2\pi i \alpha'X_{36}} - 1}\\& + \frac{1}{e^{2\pi i \alpha'X_{38}} - 1}
    + \frac{1}{e^{2\pi i \alpha'X_{16}} - 1}\frac{1}{e^{2\pi i \alpha'X_{14}} - 1} + \frac{1}{e^{2\pi i \alpha'X_{58}} - 1}\frac{1}{e^{2\pi i \alpha'X_{14}} - 1}\\& + \frac{1}{e^{2\pi i \alpha'X_{16}} - 1}\frac{1}{e^{2\pi i \alpha'X_{36}} - 1}
    + \frac{1}{e^{2\pi i \alpha'X_{36}} - 1}\frac{1}{e^{2\pi i \alpha'X_{38}} - 1}\\& + \frac{1}{e^{2\pi i \alpha'X_{38}} - 1}\frac{1}{e^{2\pi i \alpha'X_{58}} - 1},
    \end{aligned}
\end{equation}

\begin{equation}
    \begin{aligned}
    &\langle{\mathcal{S}_{2,(25,27)}\otimes\mathrm{Stokes}_{2, (25,27)}(z_1,z_2),\mathcal{S}_{2,(25,27)}\otimes\mathrm{Stokes}_{2, (25,27)}(z_1,z_2)\rangle}=\\
    & 1 + \frac{1}{e^{2\pi i\alpha' X_{27}} - 1}  + \frac{1}{e^{2\pi i \alpha'X_{25}} - 1} + \frac{1}{e^{2\pi i\alpha' X_{16}} - 1}+ \frac{1}{e^{2\pi i \alpha'X_{27}} - 1}\\& + \frac{1}{e^{2\pi i \alpha'X_{14}} - 1}
    + \frac{1}{e^{2\pi i \alpha'X_{27}} - 1}\frac{1}{e^{2\pi i \alpha'X_{25}} - 1} + \frac{1}{e^{2\pi i \alpha'X_{16}} - 1}\frac{1}{e^{2\pi i \alpha'X_{25}} - 1}\\& + \frac{1}{e^{2\pi i \alpha'X_{27}} - 1}\frac{1}{e^{2\pi i \alpha'X_{47}} - 1}
    + \frac{1}{e^{2\pi i \alpha'X_{47}} - 1}\frac{1}{e^{2\pi i \alpha'X_{14}} - 1}\\& + \frac{1}{e^{2\pi i \alpha'X_{14}} - 1}\frac{1}{e^{2\pi i \alpha'X_{16}} - 1},
    \end{aligned}
\end{equation}

and

\begin{equation}
    \begin{aligned}
    &\langle{\mathcal{S}_{2,(36,38)}\otimes\mathrm{Stokes}_{2, (36,38)}(z_1,z_2),\mathcal{S}_{2,(36,38)}\otimes\mathrm{Stokes}_{2, (36,38)}(z_1,z_2)\rangle}=\\
    & 1 + \frac{1}{e^{2\pi i\alpha' X_{38}} - 1}  + \frac{1}{e^{2\pi i \alpha'X_{36}} - 1} + \frac{1}{e^{2\pi i\alpha' X_{27}} - 1}+ \frac{1}{e^{2\pi i \alpha'X_{38}} - 1}\\& + \frac{1}{e^{2\pi i \alpha'X_{25}} - 1}
    + \frac{1}{e^{2\pi i \alpha'X_{38}} - 1}\frac{1}{e^{2\pi i \alpha'X_{36}} - 1} + \frac{1}{e^{2\pi i \alpha'X_{27}} - 1}\frac{1}{e^{2\pi i \alpha'X_{36}} - 1}\\& + \frac{1}{e^{2\pi i \alpha'X_{38}} - 1}\frac{1}{e^{2\pi i \alpha'X_{58}} - 1}
    + \frac{1}{e^{2\pi i \alpha'X_{58}} - 1}\frac{1}{e^{2\pi i \alpha'X_{25}} - 1}\\& + \frac{1}{e^{2\pi i \alpha'X_{25}} - 1}\frac{1}{e^{2\pi i \alpha'X_{27}} - 1}.
    \end{aligned}
\end{equation}

Now, squares come with weight $\frac{2}{6}$ and pentagons with weight $\frac{1}{6}$. It may now be readily verified that the weighted sum of the preceding intersection numbers yields the schematic expression,

\begin{equation}
    \frac{16}{6} + \frac{3}{2}\left(\sum_{(ij)} \frac{1}{e^{2\pi i \alpha' X_{ij}}-1}\right) + \left(\sum_{(ij,kl)} \frac{1}{e^{2\pi i \alpha' X_{ij}}-1}\frac{1}{e^{2\pi i \alpha' X_{kl}}-1}\right),
\end{equation}

where $(ij)$ is a partial quadrangulation in the first sum and $(ij,kl)$ is a quadrangulation in the second.

\end{document}